\documentclass{webofc}
\usepackage[varg]{txfonts}   
\begin{document}
\title{Search for Deeply Bound Kaonic Nuclear States with AMADEUS}
%
%

\author{\firstname{Magdalena} \lastname{Skurzok}\inst{1}\fnsep\thanks{\email{magdalena.skurzok@uj.edu.pl}} \and
\firstname{Michael} \lastname{Cargnelli}\inst{2} 
\and
\firstname{Catalina} \lastname{Curceanu}\inst{3}
\and
\firstname{Raffaele} \lastname{Del Grande}\inst{3}
\and
\firstname{Laura} \lastname{Fabbietti}\inst{4,5}
\and
\firstname{Carlo} \lastname{Guaraldo}\inst{3}
\and
\firstname{Johann} \lastname{Marton}\inst{2}
\and
\firstname{Pawel} \lastname{Moskal}\inst{1}    
\and
\firstname{Kristian} \lastname{Piscicchia}\inst{6,3} 
\and
\firstname{Alessandro} \lastname{Scordo}\inst{3}            
\and
\firstname{Michal} \lastname{Silarski}\inst{1}            
\and
\firstname{Diana Laura} \lastname{Sirghi}\inst{3}  
\and
\firstname{Ivana} \lastname{Tucakovic}\inst{7} 
\and
\firstname{Oton} \lastname{Vazquez Doce}\inst{4,5} 
\and
\firstname{S{\l}awomir} \lastname{Wycech}\inst{8} 
\and
\firstname{Eberhard} \lastname{Widmann}\inst{2} 
\and
\firstname{Johann} \lastname{Zmeskal}\inst{2} 
\
}

\institute{Institute of Physics, Jagiellonian University, 30-059 Krakow, Poland
\and
		   Stefan-Meyer-Institut f\"ur subatomare Physik, Vienna, Austria
\and
           INFN, Laboratori Nazionali di Frascati, 00044 Frascati, Italy            
\and
           Excellence Cluster "Origin and Structure of the Universe", 85748 Garching, Germany 
\and
           Physik Department E12, Technische Universit{\"a}t M{\"u}nchen, 85748 Garching, Germany       \and
           CENTRO FERMI - Museo Storico della Fisica e Centro Studi e Ricerche "Enrico Fermi", Roma, Italy   \and
           Ruger Boskovi{\'c} Institute, Zagreb, Croatia
\and
           National Centre for Nuclear Research, 00681 Warsaw, Poland
          }

\abstract{
We briefly report on the search for Deeply Bound Kaonic Nuclear States with AMADEUS in the $\Sigma^{0} p$ channel following $K^{-}$ absorption on $^{12}\hspace{-0.03cm}\mbox{C}$ and outline future perspectives for this work.

}
\maketitle
\section{Introduction}
\label{intro}

The existence or not of the Deeply Bound Kaonic Nuclear States (DBKNS) is, currently, one of the hottest topics in nuclear and
hadronic strangeness physics, from both experimental and theoretical points of view. The existence
of bound kaonic nuclear states, also called kaonic nuclear clusters, was first predicted
in 1986~\cite{Wycech}. The mass of the $\Lambda(1405)$ reflects the strength of the $\bar{\mathrm{K}}$N interaction, thus influencing the possible formation of $\bar{\mathrm{K}}$ multi-nucleon bound states. Recent theoretical calculations, based on different approaches, deliver a wide range of binding energies and widths for the di-baryonic kaonic bound state ppK$^-$~\cite{Yamazaki,Wycech2,Barnea,Oset,Dote,Ikeda} while experimental results remain contradictory \cite{plb7,plb8,plb9,plb10,mas11,mas12,mas13}.
Moreover, the extraction of a ppK$^-$ signal in K$^-$ absorption experiments is strongly affected by the dynamics of the competing K$^-$ multi-nucleon absorption processes.~Therefore, in order to clarify this issue, experimental data are needed. This research is very important in understanding the fundamental laws of Nature and the Universe and can have important consequences in various sectors of physics, such as nuclear and particle physics, as well as astrophysics. The binding of the kaon in the nuclear medium may impact on models describing the structure of neutron stars through their Equation of State~\cite{Nelson,Scordo} and collapsing binaries which are now known to be sources of gravitational waves~\cite{Abbott}.~Investigation of stable forms of strange matter like DBKNS in extreme conditions would be helpful for a better understanding of elementary kaon-nucleon interaction for low energies in non-perturbative quantum chromodynamics (QCD) and would contribute to solving crucial problems in hadron physics: hadron masses (related to the chiral symmetry breaking), hadron interactions in the nuclear medium and the structure of the dense nuclear matter.

The AMADEUS group has developed a method with a high probability for the discovery of DBKNS corresponding to $K^{-}$pp , $K^{-}$ppn and $K^{-}$ppnn kaonic nuclear clusters and their decays to $\Sigma^{0}$/$\Lambda$p, $\Sigma^{0}$/$\Lambda$d and $\Sigma^{0}$/$\Lambda$t, respectively. The method is based on the exclusive measurement of the momentum, angular and invariant mass spectra for correlated $\Sigma^{0}$/$\Lambda$p, $\Sigma^{0}$/$\Lambda$d, $\Sigma^{0}$/$\Lambda$t pairs~\cite{Curceanu}. Possible DBKNS may be produced with $K^{-}$ stopped in helium or carbon and then decaying into the considered decay channels. 
The AMADEUS experiment is conceived to perform a systematic investigation of the low-energy $\bar{\mathrm{K}}$N interaction, taking advantage of the kaons beams delivered by the DA$\Phi$NE collider \cite{dafne}. The ongoing AMADEUS analyses refer to two data samples. One is represented by the data collected by the KLOE \cite{Bossi} collaboration during the 2004/2005 data taking, corresponding to $\sim$ 1.74 fb$^{-1}$. The KLOE detector \cite{kloedet} is used as an active target, with the hadronic interaction of negative kaons with the materials of the apparatus being investigated; in particular K$^-{}$-$^9$Be absorptions in the DA$\Phi$NE thin Beryllium cylindrical layer and the DA$\Phi$NE aluminised beryllium pipe, K$^-{}$-$^{12}$C and K$^-$-H absorptions in the KLOE Drift Chamber (DC) inner wall (aluminised carbon fiber)~\cite{kloedc}, and K$^-{}$-$^4$He in the DC gas.
Extremely rich experimental information is contained in this sample, with K$^-$ hadronic captures both at rest and in-flight~\cite{piscicchiabormio}. In order to increase the statistics, and as an essential interpretation tool, a high purity carbon target (graphite) was produced in summer 2012 and installed inside the KLOE detector, between the beam pipe and the DC inner wall. The geometry of the target was optimized to maximize the kaon stopping power. The total collected integrated luminosity was $\sim$90 pb$^{-1}$.~Up till now sample of 37 pb$^{-1}$ reconstructed data was analysed. The adopted procedure is to exclusively measure the correlated $\Sigma^{0}$/$\Lambda$p, $\Sigma^{0}$/$\Lambda$d and $\Sigma^{0}$/$\Lambda$t pairs, searching for $K^{-}$pp , $K^{-}$ppn and $K^{-}$ppnn multi-nucleon, and related bound states, production.  In the next section we present briefly results of data analysis devoted to the $\Sigma^{0}$p channel. A more detailed description can be found in Ref.~\cite{Oton}.

\section{$\Sigma^{0}$p data analysis}\label{sec-1}

\begin{figure}[h!]
\centering
\includegraphics[width=13.0cm,height=3.5cm]{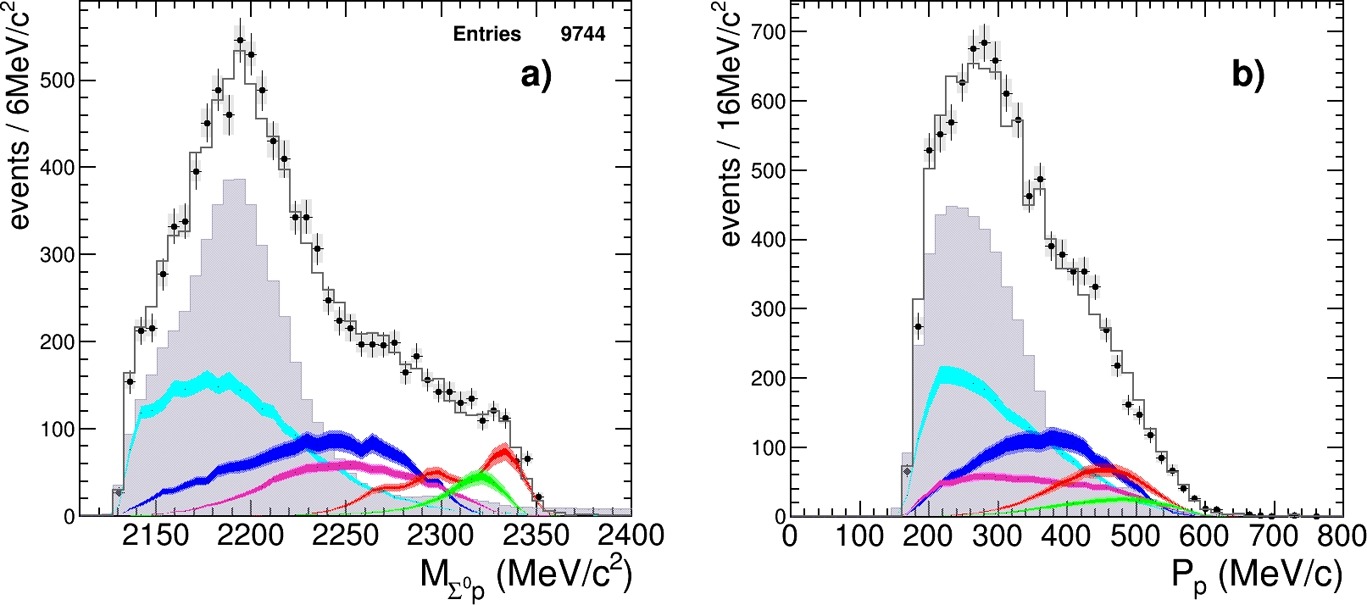}
\vspace{-0.3cm}
\caption{The simultaneous fit to the $\Sigma^{0} p$ invariant mass (left) and proton momentum distributions (right) performed for kaons stopped in DC wall. The colors represent the following fit contributions: $\pi^{0}$ background (grey), 4NA with uncorrelated production of the $\Sigma^{0} p$ final state (cyan), 3NA (blue), 2NA-FSI (magenta) and 2NA-QF (red). The black line shows the overall fit. Figure is adopted from~\cite{Oton}.\label{Lambda_t_inv}}
\end{figure}

The analysis presented below is focused on studies of $K^{-}$ absorption processes inside the Drift Chamber (DC) carbon wall leading to $\Sigma^{0}$p final states~\cite{Oton}. Selection of $\Lambda$(1116) hyperons being the signature of the $K^{-}$ hadronic interaction in $^{12}\hspace{-0.03cm}\mbox{C}$ was the first aim of the data analysis.~$\Lambda$s were identified by reconstruction of their decay vertex $\Lambda \rightarrow p + \pi^{-}$ (BR = 63.8\%). The protons and negatively charged pions selection and identification was carried out using the dE/dx information in DC installed around the interaction point and the track momenta.~The proton-pion invariant mass was determined under the $p$ and $\pi^{-}$ mass hypothesis. The constructed $p\pi^{-}$ invariant mass resulted in a value of 1115.753$\pm$0.002 MeV/c$^2$ and a statistical error of 0.5 MeV/c$^2$. Then, $\Lambda$-proton vertices were searched for and $K^{-}$ absorption processes inside the DC wall were selected based on the radial position $\rho_{\Lambda p}$.
The identification of $\Sigma^{0}$ candidates was carried out through their decay into $\Lambda \gamma$ pairs where $\gamma$ quanta were identified in the Calorimeter~\cite{kloeemc}. Afterwards, a simultaneous global fit to the $\Sigma^{0} p$ invariant mass, cos($\theta_{\Sigma^{0} p}$) and the $\Sigma^{0}$ and the proton momenta was performed in order to extract contributions from the different absorption processes like 2 Nucleon Absorption (2NA),  with final state interaction (FSI) or without (QF), 3NA and 4NA. For this purpose Monte Carlo simulations of the absorption processes in $^{12}\hspace{-0.03cm}\mbox{C}$ have been carried out. The final fit delivered the contribution of the different absorption processes. The fit results in the first measurement of the 2NA free from FSI which was found to be 12\% of the total absorption cross-section. 
The contribution from the $K^{-}pp$ bound state was included in the second fit of the experimental data.
A Breit-Wigner shape was used to simulate the bound state contribution, scanning a grid of possible bindings (15-75 MeV/c$^{2}$) and widths (30-70 MeV/c$^{2}$). The best value of the total reduced $\chi^{2}$ was achieved for the hypothesis of a binding energy of 45 MeV/c$^{2}$ and width 30 MeV/c$^{2}$ and are shown in Fig.~\ref{Lambda_t_inv}. The $K^{-}pp$ yield extracted from the fit is $K^{-}pp/K^{-}_{stop}$=(0.044$\pm$0.009$stat^{+0.004}_{-0.005}{syst}$)$\cdot$10$^{-2}$. However the 1$\sigma$ significance of the result doesn't allow the observation of the bound state to be claimed.  


\section{Summary and conclusion}
\label{sec-2}

One of the AMADEUS's goals is to search for Deeply Bound Kaonic Nuclear States by studies of $K^{-}$ absorption processes in various targets. In this work investigation of $\Sigma^{0} p$ final state was presented. It results in the extraction of contributions from various few nucleon absorption processes. The contribution from a possible $K^{-}pp$ bound state was determined, however the significance of the result is not sufficient to claim the observation of a bound system.
Presently, the analysis of the channels like $\Lambda p$, $\Lambda d$ and $\Lambda t$ is in progress. Moreover, a feasibility study \cite{bazzi_nim,bazzi_jou} is ongoing for the realization of a dedicated AMADEUS experimental setup, in order to deepen and extend the low energy anti-kaon nuclei interaction studies and obtain fundamental input for the understanding of the QCD with strangeness. 

\section{Acknowledgements}
\label{sec_ackn}

\small
We acknowledge the KLOE Collaboration for their support and
for having provided us with the data and the tools to perform the analysis presented in this paper.
We also thank as well the DA$\Phi$NE  staff for the excellent 
working  conditions  and  permanent  support. 
We acknowledge the CENTRO FERMI - Museo Storico della Fisica e Centro Studi e Ricerche 'Enrico Fermi', for the progect PAMQ.
Part  of  this  work  was  supported  by  the  Austrian   Science   Fund   (FWF):   [P24756-N20]; 
Austrian  Federal  Ministry  of  Science  and  Research  BMBWK  650962/0001  VI/2/2009;  
 the  Grant{\"i}n-Aid for Specially Promoted Research (20002003); Minstero degli Affari Esteri e della Cooperazione Internazionale, Direzione Generale per la Promozione del Sistema Paese (MAECI), Strange Matter project; Polish National Science Center, grant No. UMO-2016/21/D/ST2/01155.

%
%

\end{document}